\documentclass[10pt]{article}
\usepackage{amssymb}
\usepackage{amstext}
\usepackage{amsmath}
\newcommand*\de{\mathrm{d}}
 
\renewcommand*\epsilon{\varepsilon}
\renewcommand*\phi{\varphi}
\renewcommand*\theta{\vartheta}
\begin{document}
  
\title{\bf Poisson Bracket for Fermion Fields: Correspondence Principle, Second Class Constraints and Hamilton-Jacobi equation } 

\author{M. Leclerc}  
\date{November 18, 2012}
\maketitle 
\begin{abstract}
We introduce a symmetric Poisson bracket that allows us to describe anticommuting fields on a classical level 
in the same way as commuting fields, 
without the use of Grassmann 
variables. By means of a simple example, we show how the Dirac bracket for the elimination of the second class constraints can be introduced,  how the classical Hamiltonian 
equations can be derived and how quantization can be achieved through a direct correspondence principle. Finally, we  show that the semiclassical limit 
of the corresponding Schr\"odinger equation leads back to the Hamilton-Jacobi equation of the classical theory. Summarizing, it is shown that  
the relations between classical and quantum theory are valid for fermionic fields in exactly the same way as in the bosonic case, and that 
there is no need to introduce anticommuting variables on a classical level.     
\end{abstract}

\section{Poisson bracket and Hamiltonian equations}\label{bracket}

Symmetric Poisson brackets have been considered in the past, for instance in \cite{droz} and \cite{franke}. As we will argue in the 
following sections, the framework developed by those authors does not seem to be applicable to the interesting cases of theories of Dirac type form. 
Our basic observation is that the transition of the fundamental bracket from bosonic to fermionic theories 
can be formally performed by the transition $[x,p] \rightarrow
\{x, p\}$, where $[\ , \ ]$ is the antisymmetric Poisson bracket and $\{\ , \ \}$ the symmetric bracket. It can be performed equally well, however, 
by the transition  $[p,x] \rightarrow \{p, x\}$, which, in view of the antisymmetry of the Poisson bracket, leads to inequivalent results. As it 
turns out, we have to consider both kind of transitions in order to get a consistent theory. This is the main difference between our approach and 
previous ones. 

We consider fermion equations that are linear in the field variables. Hermitian Lagrangians are then constructed with terms of the form 
$\bar\psi_1\psi_2$ or $\bar\psi\psi$, where the variations with respect to  $\psi$ and $\bar\psi$ are carried out independently. 
Our starting point is therefore the observation that fermionic variables always occur in pairs whose elements have conjugate 
transformation behavior, e.g., with respect to $U(1)$ or Lorentz transformation. We say that the first element is of \textit{type I} and 
the second one of \textit{type II}. In addition, we claim that the momentum variable corresponding to a variable of type I is of type II and 
vice versa. In Dirac theory for instance, $\psi$ is of type I, the corresponding momentum $\pi = \delta \mathcal L / \delta \dot \psi$ of type II, while 
$\bar\psi$ is of type II and $\bar\pi = \delta \mathcal L / \delta  \dot{\bar\psi}$ of type I again.

A similar kind of variable pairs can occur in a bosonic theory, e.g., $\mathcal L = \frac{1}{2}\phi_{,m}\phi^{*}_{,m} -\frac{1}{2} m^2 \phi\phi^*$. As we 
will see, there is no need to differentiate two types in that case. 

For simplicity, we introduce the Poisson brackets for one pair of (point) variables $x_1$ (type I) and $x_2$ (type II). Consider two observables $A,B$ 
depending on both variables and their corresponding momenta, $A = A(x_1, x_2, p_1, p_2)$, $B = B(x_1, x_2, p_1, p_2)$. 

If $x_1$ and $x_2$ are fermionic\footnote{Actually, in our description, there is no difference at all between fermionic and bosonic matter, as 
long as we are dealing with the classical theory. We are simply re-expressing the same theory in terms of a symmetric bracket instead of the  
 antisymmetric Poisson bracket, in order to be able to use a direct correspondence principle between the symmetric bracket and the operator  
anticommutator for fermionic fields once the theory is to be quantized.}, we define
\begin{equation}\label{def1}
\{ A, B \} =  i \left( \frac{\partial A}{\partial x_1}\frac{\partial B}{\partial p_1}  - \frac{\partial A}{\partial x_2}\frac{\partial B}{\partial p_2} \right)
+\  i \left( \frac{\partial A}{\partial p_1}\frac{\partial B}{\partial x_1} -   \frac{\partial A}{\partial p_2}\frac{\partial B}{\partial x_2} \right).
\end{equation}
Similarly, in the bosonic case, we define 
\begin{equation}\label{def2}
[ A, B ] =  i \left( \frac{\partial A}{\partial x_1}\frac{\partial B}{\partial p_1}  + \frac{\partial A}{\partial x_2}\frac{\partial B}{\partial p_2} \right)
-\ i  \left( \frac{\partial A}{\partial p_1}\frac{\partial B}{\partial x_1} +   \frac{\partial A}{\partial p_2}\frac{\partial B}{\partial x_2} \right).
\end{equation}
The bosonic case is the standard Poisson bracket for commuting matter. The factor $i$ is introduced for convenience, to get a more direct correspondence 
to the quantum theory. Note that in the bosonic case, $x_1$ and $x_2$ enter the bracket in a symmetric way. 

The fermionic case is characterized by its symmetry in $x_1$ and $p_1$ as well as in $x_2$ and $p_2$, which will lead ultimately to the anticommutator of 
quantum mechanics. In addition, the type I variable enter in a different way as the type II ones. The fundamental brackets are found in the form
\begin{eqnarray}\label{fundbrack1}
\{ x_1, x_1\} = 0, \ \ & \{ p_1, p_1\} = 0,\ \ & \{ x_1, p_1\} = \{p_1,x_1\} =  i, \\ \label{fundbrack2}
\{ x_2, x_2\} = 0, \ \ &\{ p_2, p_2\} = 0,\ \  & \{ x_2, p_2\} = \{p_2,x_2\} = -i.
\end{eqnarray}
Both (\ref{fundbrack1}) and (\ref{fundbrack2}) can be viewed as symmetric counterparts of the antisymmetric case, 
depending on whether we start from $[x,p] = i$ or from $[p,x] = -i$. The symmetric bracket for type I variables has been introduced in \cite{droz}, 
see also \cite{franke}. We will show in the next section that the symmetric bracket cannot be used without the distinction between the two 
types of variables for Dirac type Lagrangians. 

The Hamiltonian equations for a Lagrangian $L(x_1, x_2)$ can be derived from the variation of $L = p_1 \dot x_1 + p_2 \dot x_2 - H(x_1, x_2, p_1, p_2)$. 
Since we are dealing with normal (commuting) variables, there is nothing special to take care of, and we find the canonical Hamiltonian equations 
in the form 
\begin{equation}\label{hameqs}
\dot p_{1,2} = - \frac{\partial H}{\partial x_{1,2}} \ \ \ and \ \ \  \dot x_{1,2} =  \frac{\partial H}{\partial p_{1,2}}. 
\end{equation}

Using these in (\ref{fundbrack1}) and (\ref{fundbrack2}), we find 
\begin{equation}\label{eqmot}
\begin{aligned} 
\{ H, x_1 \}\ =\ \{x_1, H\} &  =  & \ \ i\ \dot x_1,   \\
\{ H, p_1 \}\ =\ \{p_1, H\} &  =  &  - i\ \dot p_1,  \\
\{ H, x_2 \}\ =\ \{x_2, H\} &  =  & - i\ \dot x_2,   \\
\{ H, p_1 \}\ =\ \{p_2, H\} &  =  & \ \ i\ \dot p_2.  
\end{aligned} 
\end{equation}
Again this can be viewed as the fermionic counterpart of $[x_1, H] = i \dot x_1$ and $[H, x_2] = -i \dot x_2$. Quite generally, for a function depending only 
on type I variables, $f(x_1, p_2)$, we find $\{ H, f\} = i \dot f$ while for type II functions $g(x_2, p_1)$, we have $\{ H, g\} = - i \dot g$.

Quantization is now performed formally by replacing the Poisson bracket (\ref{def1}) by the operator anticommutator in the fermionic case, 
and similarly the bracket (\ref{def2}) by the commutator in the bosonic case. However, 
since realistic fermion Lagrangians are linear in the velocities, second class constraints will arise and the 
 relations derived in this section cannot be retained in the above form. 

In the next section, we consider a concrete example and show how to deal with the constraints of the theory by means of the Dirac brackets. Once this has been 
done, it is possible to pass on to the quantization of the theory.      

\section{Second class constraints, Dirac brackets and quantization}\label{secconst}

We will not repeat in detail the Dirac procedure for dealing with second class constraints, but refer instead to the original work \cite{dirac}.
The main result is that second class constraints, i.e., constraints with non-vanishing Poisson bracket, 
 arise because there are non-physical variables  in the theory that have to be 
removed consistently before the transition to the quantum theory is possible. Dirac shows that, if we have 
second class constraints $\phi_n \approx 0$ with $[\phi_n, \phi_m] = \omega_{nm}$, the consistent way 
to remove the non-physical variables is to introduce the so-called Dirac bracket
\begin{equation}\label{diracbacketdef}
[ A, B ]_D = [A, B] \ -\  [A, \phi_m]\ (\omega^{-1})_{mn}\ [\phi_n, B], 
\end{equation}
where $ \omega^{-1}$ is the inverse matrix of $\omega_{mn}$ and summation over $m,n$ is understood. The notation $\phi \approx 0$ 
is a so called weak equality, meaning that it holds only on the constraint surface. 
After the replacement of the Poisson bracket by the Dirac bracket, the constraints can be used to remove 
the non-physical variables explicitly, i.e., strong relations $\phi = 0$ can be imposed. Quantization can then be performed 
using the correspondence principle between the Dirac bracket and the commutator. 

In quite the same manner, we will introduce the Dirac bracket corresponding to the symmetric Poisson bracket
 \begin{equation}\label{diracbacket}
\{ A, B \}_D = \{A, B\} \ -\  \{A, \phi_m\}\ (\tilde\omega^{-1})_{mn}\ \{\phi_n, B\},   
\end{equation}
where $\tilde \omega_{mn} = \{\phi_m, \phi_n\}$. Note that the symmetry property $\{A, B\}_D = \{B, A\}_D$ 
also holds for the new bracket. 

Instead of a general consideration, we consider a concrete Lagrangian for a point particle, whose structure is formally similar to the structure of the 
Dirac Lagrangian,
\begin{equation} \label{lagrangian}
L = i(\bar z \dot z -  \dot{\bar z} z) - 2 \bar z z,   
\end{equation} 
where $\bar z = z^*$ is the complex conjugate of $z$. Variation yields the equations of motion
\begin{equation}\label{lagrangeEqs}
\dot{\bar z} = i \bar z \ \ \ \mathrm{and} \ \ \ \dot z = - i z.
\end{equation}
The canonical momenta $p = i \bar z$ and $\bar p = - i z$ give rise to the following constraints
\begin{equation}\label{constraints}
\phi_1 = p - i\bar z \approx 0 \ \ \ \mathrm{and}\ \ \  \phi_2  = \bar p + i z \approx 0. 
\end{equation}
With  $z$ beeing of type I (and thus, $p$ of type II) and $\bar z$ of type II (and $\bar p$ of type I), we find with the help of (\ref{fundbrack1})
and (\ref{fundbrack2})
\begin{equation}\label{constrcommutator}
\{\phi_1, \phi_2\} = \{\phi_2, \phi_1\} = i \{p,z\} - i \{\bar z , \bar p\} = - 2, 
\end{equation}
which means that $\phi_1$ and $\phi_2$ are second class constraints. Note that the antisymmetric 
Poisson bracket (\ref{def2}) results in similar relations. On the other hand, if both $z$ and $\bar z$ are 
considered to be of type I, i.e., with the symmetric bracket as defined in \cite{droz} and \cite{franke}, then we end up    
with commuting constraints, i.e., 4 independent variables remain in the Hamiltonian theory. 
It is not hard to show that this will not lead to a consistent theory on the quantum level.  

Straightforward application of the 
Dirac procedure (\ref{diracbacket}) leads to 
\begin{equation} \label{comrelations}
\{ z, p \}_D = \frac{i}{2}, \ \ \{z, z \}_D= 0, \ \ \{p,p\}_D = 0,
\end{equation}
while all other Dirac brackets containing  $\bar z$ and $\bar p$ can be deduced by inserting 
the variables into (\ref{comrelations}) using the constraints (\ref{constraints}) as strong relations.
Straightforward definition of $H = p \dot z + \dot{\bar z} \bar p - L$ and strong imposition of the constraints leads to the Hamiltonian
\begin{equation}\label{hamiltonian}
H = - 2 i p z. 
\end{equation}
Note that the Hamiltonian equations in the form (\ref{hameqs}) are not valid neither for $p$ nore for $z$. This is a result of the second class constraints, the 
same would have happened with the classical antisymmetric Poisson brackets. On the other hand, the relations (\ref{eqmot})
\begin{equation}\label{eqmot2}
\begin{aligned} 
\{ H, z \}_D \ =\ \{z, H\}_D &  =  & \ \ i\ \dot z,   \\
\{ H, p \}_D \ =\ \{p, H\}_D &  =  & \ \  - i\ \dot p,  
\end{aligned} 
\end{equation}
are still valid, it is sufficient to replace the Poisson bracket with the corresponding Dirac bracket. They lead to the Hamiltonian equations 
\begin{equation}\label{hameqs2}
z = i \dot z, \ \ \ \mathrm{and}\ \ \  p = - i\dot p, 
\end{equation} 
which upon imposing the constraints are equivalent to  (\ref{lagrangeEqs}).  

Following Dirac \cite{dirac}, we perform quantization by replacing the symmetric Dirac bracket by the operator anticommutator 
\begin{equation}\label{quant}
\{ A, B \}_D \ \rightarrow \  \{A, B\}_Q = AB + BA.
\end{equation}

In the next section, we will introduce a concrete representation that leads us to the Schr\"odinger formulation of the theory. We close this section by 
pointing out the important difference between the equations of motion and the generator of time translations. Let again $f = f(z, \bar p)$ be a function 
of type I variables and $g = g(\bar z, p)$ a function of type II variables. The equations of motion then read
\begin{equation}\label{eqmota}
\begin{aligned} 
\{f, H\}_D &  =  & \ \ i\ \dot f,   \\
\{g, H\}_D &  =  & \ \  - i\ \dot g,  
\end{aligned} 
\end{equation}
while no general statement can be made concerning the bracket of $H$ with a mixed function, e.g., $h(z, p)$. In the quantized theory, 
the operator relations are of the same form:
\begin{equation}\label{eqmotb}
\begin{aligned} 
\{f, H\}_Q &  =  & \ \ i\ \dot f,   \\
\{g, H\}_Q &  =  & \ \  - i\ \dot g,  
\end{aligned} 
\end{equation}
where $\{\ ,\ \}_Q$ is now the anticommutator. Again, we don't know anything about the anticommutators of $H$ with mixed functions. For instance, you can 
evaluate $\{H, H\}_Q$, it will be different from zero, although $H$ is actually a constant of motion.

On the other hand, the commutator in quantum mechanics, $[A,B]_Q = AB - BA$, can be shown to be the generator of time translations. Indeed, for 
$f = f(z, \bar p)$ a type I and $g = g(\bar z, p)$ a type II function, we find, e.g., for the product $gf$
\begin{eqnarray*}
[gf, H]_Q &= &g[f,H]_Q + [g,H]_Q f = gfH - gHf + gHf - Hgf \\
&=& g \{f, H\}_Q - \{g,H\}_Q f =   i  g \dot f  + i \dot g f \\
&=& i \frac{\de }{\de t} (gf)  , 
\end{eqnarray*}
where the equations of motions have been used in the second line. Quite generally, we have in the quantum theory 
\begin{equation} \label{timetransquant}
[h,H]_Q =  i\dot h
\end{equation}
for 
any function $h$. This holds irrespectively whether the theory has been quantized with commutators or with anticommutators. For this reason, 
even in the fermionic case,  
the equations of motion are often written in the commutator form $[q, H]_Q = i \dot q$ and $[p, H]_Q =  i \dot p$, arguing that $H$ is 
bilinear in the field variables or similar, see, e.g., \cite{pauli}. In our opinion, this line of argumentation somehow 
undermines the correspondence principle, 
which is be based on anticommutators in the fermionic case, while the fact that $H$ is the generator of time translations is an additional 
property that has to be shown separately.

It is not possible to deduce the same relation in  the classical theory. This is again a result of the second class constraints. Only in those cases
where the Dirac bracket is identical with the Poisson bracket, a direct evaluation of $[gf,H]$ with (\ref{def1}) and (\ref{def2}), using 
$[f,H ] =  \{f,H\}$ and $[g,H]  = - \{g, H\}$,  leads  to $[gf,H] =   i \frac{\de }{\de t} (gf)$. This is not true in systems with second class constraints, 
like the example considered in this section. This is simply because the Hamiltonian equations are not valid in their canonical form.

The reason for this discrepancy between classical and quantum case is that 
the correspondence principle (\ref{quant}) is based on the (symmetric) Dirac bracket and the anticommutator. It is designed to reproduce the correct 
equations of motion. There is no correspondence between the commutator and the (antisymmetric) Poisson bracket. 

In the bosonic case, things are different. From the antisymmetric Poisson bracket, we get an antisymmetric Dirac bracket, which is related via the 
correspondence principle to the commutator in quantum mechanics. Classically, we have $[f, H]_D =  i \dot f $ and $[g, H]_D =  i \dot g$, 
thus there is no difference between type I and type II. Clearly, we also have $[fg, H]_D =  i \frac{\de }{\de t} (fg)$ and 
the corresponding relation $[fg, H]_Q =  i \frac{\de }{\de t} (fg)$ in the quantum mechanical 
case. So, incidentally, the generator of time translations can be represented by the same bracket structure as the equations of motion. 

It is not hard to find the classical relation for time translations in the fermionic theory. Since classically, their is no difference between the 
fermionic and the bosonic theory, we can still use the bosonic Dirac bracket as it is defined in (\ref{diracbacket}), and find $[h, H]_D =  i \dot h$. 
However, we should then be aware that there is no correspondence between $[\ ,\ ]_D$ and the commutator $[\ ,\ ]_Q$ in the fermionic case.

Alternatively, one can directly try to find a decent expression for the time translations for the specific theory in question. 
In our example, if we consider functions $h(z,p)$ (the remaining variables can be removed anyway with the help of the constraints), then we can show
\begin{eqnarray*}
i \dot h &=& i \frac{\partial h}{\partial z} \dot z  + i \frac{\partial h}{\partial p} \dot p \\
&=& \{z,H\}_D\  \frac{\partial h}{\partial z} - \frac{\partial h}{\partial p} \ \{p, H\}_D \\
&=& \{z, p\}_D \ \frac{\partial h}{\partial z}\frac{\partial H}{\partial p} - \{x,z\}_D \ \frac{\partial h}{\partial p}\frac{\partial H}{\partial z}, 
\end{eqnarray*}
where the equations of motion have been used. Comparing with(\ref{def1}), we find
\begin{equation}\label{timetransclass}
 i \dot h  = - i\{z,p\}_D\ [h, H]. 
\end{equation}
The results of this section can easily be generalized to functions with a direct time dependence.

\section{Hamilton-Jacobi equation and Schr\"odinger representation}\label{schroedinger}

We start with a few preliminary remarks concerning the secondary constraints. For simplicity, we review the bosonic case first. 

There are several ways to deal with second class constraints in the Hamiltonian formalism, see, e.g., 
\cite{gueler}, \cite{rothe} or \cite{dominici}. 
A convenient way to avoid in a certain sense the constraints is the so called Faddeev-Jackiw formalism \cite{faddeev}, 
which essentially 
consists of re-adjusting the Lagrangian (with the help of surface terms) in a way to put it into the first order form 
$L = p\dot q - H$. The remaining variables (not appearing in the kinetic term $p\dot q$) can be considered as Lagrange multipliers.
For our example (\ref{lagrangian}), this means putting it into the form 
\begin{equation}\label{faddeev1}
L = 2 i (\bar z \dot z) - 2 \bar z z.
\end{equation}
The momentum  $\pi = 2 i \bar z$ can be directly read off and the Hamiltonian is simply $ 2 \bar z z = - i \pi z$.
No constraints arise explicitly in this formalism\footnote{In fact, the constraints are still there, namely $\bar \pi = 0$ and 
$\pi = 2 i \bar z$, and they are still second class, 
but the choice of variables is now such that the dynamical variables $z$ and $\pi$ satisfy canonical Dirac brackets, i.e., identical to the Poisson 
brackets, and the remaining variables can be ignored.} and the canonical relation $[z ,\pi] = i$ (in the bosonic case) is valid. 

For such an unconstrained theory, we get the Hamilton-Jacobi equation $H = -  \frac{\partial S}{\partial t}$ from the 
replacement of the momentum $\pi$ by $\frac{\partial S}{\partial z}$ in the Hamiltonian, which leads to 
\begin{equation}\label{hjdfaddeev}
- i \frac{\partial S}{\partial z} z= -   \frac{\partial S}{\partial t}.
\end{equation} 
Here instead, we wish to work directly with the original Lagrangian and start therefore from the Hamiltonian $H = - 2 i p z$
derived in the previous section. We consider the action $S$ as a function of time and coordinates, where we assume 
that the unphysical variables have been removed by the Dirac procedure. The total time derivative of the action then reads
\begin{equation}\label{totalder}
\begin{aligned}
\frac{\de S}{\de t} = L &=& \frac{\partial S}{\partial z} \ \dot z + \frac{\partial S}{\partial t} \\
&=& \frac{\partial S}{\partial z} \ \dot z  - H, 
\end{aligned}
\end{equation}
where the Hamilton-Jacobi equation $H = -\frac{\partial S}{\partial t}$ has been used. 
In (\ref{totalder}), we use the Hamiltonian equations of motion and write
\begin{equation}\label{lagrangian2}
\begin{aligned}
L &= - i\ \frac{\partial S}{\partial z} [z,H]_D  - H \\
&=  - i\ \frac{\partial S}{\partial z} [z,p]_D \frac{\partial H}{\partial p} - H. 
\end{aligned}
\end{equation}
On the other hand, since the Hamiltonian corresponds to an unconstrained theory, we can formally get back the 
Lagrangian via the inverse Legendre transform\footnote{In general, we get back an equivalent Lagrangian, 
differing from the original one by a surface term.} 
\begin{equation}\label{legendre}
L = p \frac{\partial H}{\partial p} - H. 
\end{equation} 
Note that on shell, this Lagrangian is zero. It is indeed not hard to show that this is true, up to a total time derivative, 
for the original Lagrangian too. If we compare  (\ref{legendre}) with (\ref{lagrangian2}), we conclude that the expression for $p$ has to be of the form
\begin{equation} \label{bosonicMom}
p  = - i [x, p]_D \ \frac{\partial S}{\partial z} = - i [S, p]_D. 
\end{equation}
Inserting this into $H(z,p) = - \frac{\partial S}{\partial t}$, we find the same Hamilton-Jacobi equation (\ref{hjdfaddeev}) 
as from the (formally) unconstrained theory. 

Now that we have found a way to deal with second class constraints, the treatment of the fermionic theory is trivial.  
We perform the transition to the Hamilton-Jacobi theory in the same manner, namely by replacing 
\begin{equation} \label{FermionicDingens}
p  = - i \{x, p\}_D \ \frac{\partial S}{\partial z} = - i \{S, p \}_D 
\end{equation}
in the Hamiltonian and write
\begin{equation}\label{HJD}
H\left(z, - i \{S, p \}_D \right) = -  \frac{\partial S}{\partial t}, 
\end{equation}
where $S = S(z,t)$.

For our specific example, we have $H = - 2 i pz $, and thus 
\begin{eqnarray*}
 H &=& -2 i pz \ =\  -  2  \{S, p\}_D\ z \\
&=&- 2 \frac{\partial S}{\partial z} \{z,p\}_D\  z 
\ =\  - 2 \frac{\partial S}{\partial z} \frac{i}{2} z \\
&=&  - i  \frac{\partial S}{\partial z} z , 
\end{eqnarray*}
leading to the Hamilton-Jacobi equation
\begin{equation}\label{hjd1}
- i  \frac{\partial S}{\partial z} z = - \frac{\partial S}{\partial t}, 
\end{equation}
which, as expected, is identical to the one obtained in the bosonic case.
 
It is not hard to solve this by the ansatz $S = - E t + F(z)$, it leads to $F = \ln(z) E i$ with constant $E$. The time dependence
of $z$ is found by setting $\partial S/\partial E = a$ with a new constant $a$. This leads to 
\begin{equation}\label{hjdsolution}
z = \exp(i(-t-a)),
\end{equation}
while the relation (\ref{FermionicDingens}) leads to 
\begin{equation}\label{hjdsolution2}
\frac{i}{2} \frac{E}{z} = p. 
\end{equation}
Equation (\ref{hjdsolution}) is the general solution of the equation of motion  $\dot z = - i z$. From (\ref{hjdsolution2}) 
it also follows $pz = \frac{i}{2} E  = const$, which, together with the time derivative of (\ref{hjdsolution}), 
leads to $\dot p =  i p$, which is the second equation in (\ref{eqmot2}). The ansatz 
(\ref{FermionicDingens}) thus leads to the correct solutions of the classical equations of motion. 

Finally, we will show that the classical theory can be obtained from the quantum theory in  a suitable limit. For this, we 
formulate the quantum theory in the Schr\"odinger representation and try to retrieve the Hamilton-Jacobi equation as the lowest 
order contribution from a semiclassical wave solution. The quantum theory is based on the correspondence between the Dirac 
bracket and the anticommutator, thus we have the following anticommutation relations (see (\ref{comrelations}))
\begin{equation} \label{comrelations2}
\{ \hat z, \hat p \}_Q = \frac{i}{2}, \ \ \{\hat z, \hat z \}_Q= 0, \ \ \{\hat p,\hat p\}_Q = 0,
\end{equation}
where we denote the quantum operators with a hat. If the left hand sides were  commutators (bosonic theory), then 
we could propose the operator representation $\hat z = z$ (multiplication) and $\hat p = -[z,p]_D \partial_z$. In the 
fermionic case, we will now use a tool that is used by many authors right from the start in the classical theory, namely 
Grassmann variables. In our case, we need only one anticommuting pair $\eta, \bar \eta$, satisfying 
\begin{equation}\label{grassmann}
\eta^2 = 0,\ \ \bar\eta^2 = 0, \ \  \eta \bar\eta = 1 = - \bar\eta\eta. 
\end{equation}
We now define the operators in the Schr\"odinger representation (in position base) by 
\begin{equation}\label{posops}
\hat z = \eta \ z, \ \ \  \hat p = -\bar\eta\ \{z,p\}_D \ \partial_z, 
\end{equation}
where the Dirac bracket $\{z,p\}_D = \frac{i}{2}$ is to be evaluated classically, i.e., by (\ref{diracbacket}). It is not hard to 
show that these operators satisfy the relations (\ref{comrelations}) and are thus suitable for our purpose. 
 
The Hamiltonion is of the form $\hat H = - 2 i \hat z \hat p$, and the Schr\"odinger equation 
$\hat H \psi(z,t) = i \partial_t \psi(z,t)$ reads
\begin{equation}\label{schroedingerequ}
-  z\ \frac{\partial }{\partial z}   \psi(z,t) =  i \frac{\partial}{\partial t}\ \psi(z,t).
\end{equation}
It is interesting to remark that we get the same Schr\"odinger equation in the bosonic theory. Differences arise 
(apart from reordering issues) as soon as we calculate expectation values of operators. 

With the semiclassical ansatz $\psi = a\exp(iS)$, where $a$ is a slowly varying function, we find for the lowest order 
contribution in $\hbar$ (recall that $S$ is actually $S/\hbar$)  
\begin{equation}\label{hjd2}
- i z \frac{\partial S}{\partial z} = - \frac{\partial S}{\partial t},
\end{equation}
which is again the Hamilton-Jacobi equation (\ref{hjd1}). It is important to remark, that in order to be able to get back the 
Hamilton-Jacobi equation in the semiclassical limit, we had to write down the Schr\"odinger equation based on the operator 
ordering $\hat H = - 2 i \hat z \hat p$. The reverse order of $\hat p$ and $\hat z$ leads to results incompatible with the 
classical limit. As opposed to this, in a bosonic theory, the operator order only influences the higher order results and cannot be 
fixed by classical arguments. 

From the Hamilton-Jacobi equation (\ref{hjd2}), without any knowledge of the original constraints, we cannot retrieve 
the original Lagrangian. The reason is that the step from the Lagrangian to the Hamiltonian is not invertible. Due to the 
constraints, it is not a genuine Legendre transform. 

Nevertheless, it is possible to find a Lagrangian equivalent to the original one. Starting from (\ref{totalder}) and 
(\ref{FermionicDingens}), we can write
\begin{equation}\label{lagrangianSimple}
L = \frac{i p}{ \{ z,p\}_D}\  \dot  z - H = \frac{1}{2} p \dot z + 2iz p. 
\end{equation}
The next step consists in replacing the momentum by the velocity using $\{z, H \}_D = i \dot z$. In our case, 
it leads to $\dot z =  - i z$, which cannot be used to replace $p$. We have to conclude that $L$ is already in its final 
form, and that $p$ plays the role of a Lagrange multiplier. Indeed, (\ref{lagrangianSimple}) is equivalent to the original 
Lagrangian (\ref{lagrangian}) if we rename the multiplier and write $i\bar z$ instead of $p$. It is actually the form 
(\ref{faddeev1}) that is used in the Faddeev-Jackiw procedure (\cite{faddeev}), where the Lagrangian is modified by  surface terms 
in order to circumvent the second class constraints.
   
\section{Discussion}

With the exception of the first section, our considerations were based on a specific, very simple, Lagrangian. The parts 
that are especially sensible to the form of the Lagrangian are the constraints and the way how to deal with them. 
However, it was not our intention to provide a general treatment of, e.g., Hamilton-Jacobi theory for constraint theories. In addition, the example is relevant enough, since it leads to constraints that are formally identical to those 
arising in Dirac theory (on a field theoretical level), and also, for instance, in Schr\"odinger theory viewed as a field theory, i.e., while performing the second quantization.  

On the other hand, the main result of this paper, namely that it is possible to describe the classical 
theory with the help of a symmetric bracket, is generally valid.  So, in a certain way, the anticommutativity of the quantum operators can be represented by a bracket structure in the classical theory. 
No Grassmann variables are needed for this purpose. On a mere classical level, it is of no importance whether a theory 
is formulated in terms of symmetric or antisymmetric brackets, or with no brackets at all. The merit, however, of 
having this symmetric bracket structure is that quantization of fermionic theories can be performed by a direct correspondence principle, in the same way as in a conventional bosonic theory.

\end{document}